\documentclass{aa}
\usepackage{graphicx, natbib,epsfig,amsmath,amssymb, mathrsfs, txfonts}
\DeclareGraphicsExtensions{.eps, .jpg, .ps}
\bibpunct{(}{)}{;}{a}{}{,}
\begin{document}

\title{Speckle temporal stability in XAO coronagraphic images \\ \textit{(Research Note)} \\  \Large{II. Refine model for quasi-static speckle temporal evolution for VLT/SPHERE}}
\author{P.\ Martinez\inst{1}, M.\ Kasper\inst{2}, A. Costille\inst{3}, J.F. Sauvage\inst{4}, K. Dohlen\inst{5}, P. Puget\inst{3}, and J.L. Beuzit\inst{3}}
\institute{Laboratoire Lagrange, UMR7293, Universit\'e de Nice Sophia-Antipolis, CNRS, Observatoire de la C\^ote d\'{}Azur, Bd. de l\'{}Observatoire, 06304 Nice, France 
\and  European Southern Observatory, Karl-Schwarzschild-Stra\ss{}e 2, D-85748, Garching, Germany
\and UJF-Grenoble 1/CNRS-INSU, Institut de Plan\'{e}tologie et d'Astrophysique de Grenoble UMR 5274, Grenoble, F-38041, France
\and Laboratoire d'Astrophysique de Marseille, UMR 7326, CNRS, Universit\'{e} de Provence, 38 rue Fr\'{e}d\'{e}ric Joliot-Curie, 13388, Marseille Cedex 13, France
\and Office National d\'{}Etudes et de Recherches Aerospatiales (ONERA), Optics Department, BP 72, F-92322 Chatillon cedex, France} 
\offprints{patrice.martinez@oca.eu}

\abstract
{Observing sequences have shown that the major noise source limitation in high-contrast imaging is due to the presence of quasi-static speckles. 
The timescale on which quasi-static speckles evolve, is determined by various factors, among others mechanical or thermal deformations.
}
{Understanding of these time-variable instrumental speckles, and especially their interaction with other aberrations, referred to as the pinning effect, is paramount for the search of faint stellar companions. The temporal evolution of quasi-static speckles is for instance required for a quantification of the gain expected when using angular differential imaging (ADI), and to determine the interval on which speckle nulling techniques must be carried out.} 
{Following an early analysis of a time series of adaptively corrected, coronagraphic images obtained in a laboratory condition with the High-Order Test bench (HOT) at ESO Headquarters, we confirm our results with new measurements carried out with the SPHERE instrument during its final test phase in Europe. The analysis of the residual speckle pattern in both direct and differential coronagraphic images enables the characterization of the temporal stability of quasi-static speckles. Data were obtained in a thermally actively controlled environment reproducing realistic conditions encountered at the telescope.}
{The temporal evolution of the quasi-static wavefront error exhibits linear power law, which can be used to model quasi-static speckle evolution in the context of forthcoming high-contrast imaging instruments, with implications for instrumentation (design, observing strategies, data reduction). Such a model can be used for instance to derive the timescale on which non-common path aberrations must be sensed and corrected.
We found in our data that quasi-static wavefront error increases with $\sim$0.7$\AA$ per minute.} 
{}

\authorrunning{Martinez et al.}
\titlerunning{Speckle temporal stability in XAO coronagraphic images II.}
\maketitle

\section{Introduction}
Observing sequences have shown that the major noise source limitation in high-contrast imaging is due to the presence of instrumental speckles, and more precisely to \textit{quasi-static} speckles \citep{Marois03, Boccaletti03, Boccaletti04, Hinkley07}. 
Speckle noise originates from wavefront errors caused by various independent sources, and evolves on different timescales pending to their nature. 
The first class of speckle to overcome comes from the large, dynamical wavefront error that the atmosphere generates, but real-time adaptive optics systems measure and correct it down to fundamental limitations.
The fast-varying speckle noise floor left uncorrected by such systems would average out over time, as it consists in a random and uncorrelated noise for which the intensity variance converges to null contribution for an infinitely long exposure. For a non-photon noise limited observing run, speckle noise associated to wavefront aberrations introduced in the optical train are fundamental to tackle. 
In this context, instrumental speckles can be divided into two different flavors:  long-timescale wavefront errors present in the optical train (e.g., optical quality, misalignment errors) that generate static speckles that constitute a deterministic contribution to the noise variance, and slowly-varying instrumental wavefront aberrations, amplitude and phase errors, originating from various causes, among others mechanical or thermal deformations.  The latest evolve on a shorter timescale than long-lived aberrations, and are the so-called quasi-static speckles. 

Instrumental speckles average to form a fixed pattern, which can be calibrated to a certain extent. A deterministic contribution to the noise variance such as static speckles can easily be calibrated, while using a reference image, time-variable quasi-static noise can be subtracted as well, but its temporal evolution ultimately limit this possibility. In particular it is understood that some timescales have a larger impact than others.
This is especially true as quasi-static speckles interact with other aberrations, referred to as the \textit{pinning effect}, or speckle cross terms.
The timescale of quasi-static speckles evolution is essential to understand and predict the performance of the new generation of instruments such as SPHERE \citep{SPHERE}, GPI \citep{GPI}, HiCIAO \citep{HiCIAO}, and Project 1640 \citep{P1640}.
The temporal evolution of these quasi-static speckles is in particular needed for the quantification of the gain expected with angular differential imaging \citep[ADI,][]{Marois2006}, as well as to determine the timescale on which speckle nulling techniques should be carried out. 
For instance, a typical hour-long ADI observing sequence provides a partial self-calibration of the residuals after a rotation of $\sim1\lambda/D$ at a given angular separation, which generally requires less than few minutes (e.g., 5 to 7 mn at 1$\arcsec$ on a 8-m class telescope for stars near the meridian in $H$-band), though it depends on wavelength, telescope latitude, and object declination. Residual speckles with decorrelation times faster than the time needed to obtain the ADI reference image cannot then be removed, while quasi-static speckles associated with larger timescales can largely be subtracted.

In this context, several authors have investigated the decorrelation timescale of quasi-static residuals in the particular context of ADI but at moderate 20-40~$\%$ Strehl levels \citep{Marois2006, 2007ApJ...660..770L, Hinkley07}, while in a former paper \citep[][hereafter Paper I]{Martinez2012}, we explored the realm of very high Strehl ratios (extreme adaptive optic  systems, XAO).
In paper I, we analyzed a time series of adaptively corrected, coronagraphic images recorded in the laboratory with the High-Order Test bench, a versatile high-contrast imaging, adaptive optics bench developed at ESO. We shown that quasi-static aberrations exhibit a linear power law with time and are interacting through the pinning effect with static speckles. 
We examined and discussed this effect using the statistical model of the noise variance in high-contrast imaging proposed by \citet{Soummer07}. 
In particular, we found that quasi-static speckles, fast-evolving on the level of a few angstroms to nanometers over a timescale of few seconds, explained the evolution of our sensitivity through amplification of the systematics. It is believed that this effect is a consequence of thermal and mechanical instabilities of the optical bench. The HOT bench is indeed localized in a classical laboratory setting, and was not initially designed to guarantee stability, nor mechanical stability at the level that would be required/expected for an actual high-contrast imaging instrument. Indeed Paper I emphasizes the importance of such stability for the next generation of high-contrast instruments, but the estimates found in this former analysis (amplitude, and slope of the temporal evolution) could not fairly be considered as representative of a realistic situation in order to affect operational aspects or designs of nowadays/future real instruments.  

In this paper, we confirm the results presented in Paper I with more representative measurements. We analyze a time series of adaptively corrected, coronagraphic images with the SPHERE instrument in a thermally actively controlled environment reproducing realistic conditions encountered at the telescope. In this context, we propose a refine model of quasi-static speckle evolution that can be used for forthcoming high-contrast imaging instrument classes that SPHERE represents.

The paper reads as follow: in Sect. 2, we briefly recall the formalism of the statistical model of the noise variance in high-contrast imaging proposed by \citet{Soummer07} and used in Paper I to discuss our former results, in Sect. 3, the experimental conditions are described, and in Sect. 4 we analyze and discuss the results. Finally, in Sect. 5, we draw conclusions. 

\section{Speckle noise and dynamical range}
In Paper I, we examined and discussed our data using the statistical model of the noise variance in high-contrast imaging proposed by \citet{Soummer07}.
Following the same formalism as for Paper I, we briefly recall here the main equations.  

\citet{Soummer07} proposed the following analytical expression for the variance of the intensity, including speckle and photon noise in the presence of static, quasi-static, and fast varying aberrations, in the context of a propagation through a coronagraph:
\begin{equation}
\sigma^{2}_I= N \left( I^{2}_{s1} + NI^{2}_{s2} + 2I_c I_{s1} + 2N I_c I_{s2} + 2 I_{s1}I_{s2} \right) + \sigma^{2}_p,
\label{variance}
\end{equation}
\noindent where $I$ denotes the intensity, $\sigma^{2}_p$ is the variance of the photon noise, and $N$ is the ratio of fast-speckle and slow-speckle lifetimes. The intensity produced by the deterministic part of the wavefront, including static aberrations, is denoted by $I_c$, while the $I_s$ terms corresponds to the halo produced by random intensity variations, i.e.\ atmospheric ($I_{s1}$) and quasi-static contributions ($I_{s2}$). 
In this generalized expression of the variance, several contributions can be identified by order of appearance: (1/) 
the atmospheric halo, (2/) the quasi-static halo, (3/) the atmospheric pinning term, the speckle pinning of the static aberrations by the fast evolving atmospheric speckles, (4/) the speckle pinning of the static by quasi-static speckles, 
and finally (5/) the speckle pinning of the atmospheric speckles by quasi-static speckles. Equation~\ref{variance} provides useful insights in the understanding of the impact of quasi-static speckles and their interactions through the pinning phenomenon with other aberrations present in a real instrument.

As for Paper I, we focus our analysis on contribution (4/) to this noise budget in XAO coronagraphic images. In particular, we are interested in the speckle pinning of the static by the quasi-static speckles when no atmospheric contribution is present (dynamical speckles), i.e., when the contribution of $I_{s1}$ to Eq.~\ref{variance} can be neglected. Furthermore, our study concerns a situation where the photon noise is not limiting, so that the contribution $\sigma^{2}_p$ from the noise variance can be neglected.
In such conditions, Eq.~\ref{variance} can be simplified such that:
\begin{equation}
\sigma^{2}_I \simeq \left(I^{2}_{s2} +  2 I_c I_{s2}  \right), 
\label{variance3}
\end{equation}
\noindent and the present study focuses on the effect of the cross-term $I_c I_{s2}$. Since we can fairly assume that $I_{s2}\ll I_c$, $I_{s2}$ can be neglected except in the cross-term, and the noise variance in the raw coronagraphic image finally becomes:
\begin{equation}
\sigma^{2}_I \simeq  2I_{c} I_{s2}. 
\label{A}
\end{equation}
Ultimately, very deep dynamic range imager, such as SPHERE, or GPI,  aim to calibrate static speckles ($I_c$) such that $I_c \approx I_{s2}$, which would largely reduce speckle pinning. 
Speckle nulling techniques to correct for remnant quasi-static aberrations would be a must to access deeper contrast level, nonetheless, the temporal characteristic of $I_{s2}$ remains a key parameter in these circumstances (Eq. \ref{variance3}). \\
From Eq. \ref{A} a breakdown of this pinning effect can be carried out at the level of differential images. 
Raw coronagraphic images are dominated by static speckle noise. This means that the interaction between the quasi-static terms of Eq.~\ref{A}, being time-dependent, and static terms, assumed time-independent, can be studied through differential imaging from a time series of raw coronagraphic images,
which simply refers to the difference in intensity between an image recorded at time $t_0 + \Delta t$ and the reference image registered at $t_0$. 
In this situation, a similar expression of the noise variance for the differential images ($\sigma_{DI}$) can be derived as the difference of Eq.~\ref{A} evaluated at $t_0 + \Delta t$, to that of the reference, at $t_0$, and reads: 
\begin{equation}
\sigma^{2}_{DI} \simeq  2I_{c}\Delta I_{s2},  
\label{B}
\end{equation}
\noindent where $\Delta I_{s2}$ represents the quasi-static evolution between the two successive images. Therefore, the quasi-static contribution can be expressed as:
\begin{equation}
\Delta I_{s2} \simeq \frac{\sigma^{2}_{DI}}{2I_{c}}.
\label{C}
\end{equation}
A general expression of the speckle intensity \citep{Racine99} is:
\begin{equation}
I_{speckle} \approx \frac{(1 - S)}{0.34},
\label{D}
\end{equation}
\noindent where S can be related to the wavefront error $\phi$ using Mar\'{e}chal's approximation \citep{BW}:
\begin{equation}
S \approx 1 - \left( \frac{2\pi \phi}{\lambda}\right)^{2},
\label{D2}
\end{equation}
and the contribution from static speckles to the wavefront error ($\phi_{s2}\ll \phi_{c}$, and $\phi_{s1}$ neglected) can be expressed as:
\begin{equation}
\phi_{c} \simeq \frac{\lambda }{\pi \sqrt{6}} \times \sqrt{I_c}.
\label{F}
\end{equation}
\noindent Similarly, the contribution from quasi-static speckles to the wavefront error in the differential images can be expressed as:
\begin{equation}
\Delta \phi_{s2} \simeq \frac{\lambda}{2\pi \sqrt6} \times \frac{\sigma_{DI}}{\sqrt I_{c}}.
\label{E}
\end{equation}
Using Eqs.~\ref{F} and \ref{E}, the analysis of both direct and differential images allows to characterize the temporal properties of static and quasi-static aberrations.   
We note that both Eq. \ref{F} and \ref{E} are approximated expressions relying on simple rule of thumbs/assumptions (e.g., general expression of speckle intensity, low phase aberration regime) to provide wavefront estimation per Fourier component.

\section{Experimental conditions}
\subsection{Optical setup}
\label{refStatic}
SPHERE which stands for Spectro-Polarimetric High-contrast Exoplanet REsearch, is a second generation instrument for the Very Large Telescope, aiming at direct detection and spectral characterization of extrasolar planets. 
The instrument is now nearing completion in its final integration stage in Europe, before shipping to Chile. Being in its final test phase, it offers a unique opportunity to carried out research on static and quasi-static speckles. SPHERE is a unique instrument, with first light in 2013, including a powerful extreme adaptive optics system (SAXO), an infrared differential imaging camera (IRDIS), an integral field spectrograph (IFS), and a visible differential polarimeter (ZIMPOL).  
The time series of adaptively corrected, coronagraphic images that will be discussed through this paper have been obtained using SPHERE with the IRDIS instrument, so that for the sake of clarity only the systems that have been used and are relevant will be further described.

The SPHERE adaptive optics for exoplanet observation (SAXO) uses a 41$\times$41 actuator deformable mirror (DM) of 180 mm diameter with inter-actuator stroke $>$ $\pm$ 1$\mu$m and a maximum stroke $>$ $\pm$ 3.5$\mu$m, and a 2-axis tip-tilt mirror (TTM) with $\pm$0.5 mas resolution. The wavefront sensor is a 40$\times$40 sub-apertures Shack-Hartmann sensor equipped with a spatial filter for aliasing minimization. During the test, no dynamical turbulence was present in the system, except the low internal turbulence (optical elements are installed inside an hermetical enclosure). Hence, SAXO is used to correct for internal turbulence, static aberrations, and guarantee image and pupil stability. As image and pupil stability are essential in high-contrast imaging, differential image movements due to thermo-mechanical effects are measured in real-time using an auxiliary NIR tip-tilt sensor located close to the coronagraph focus plan, and corrected with a differential tip-tilt mirror in the wavefront sensor arm. Similarly, pupil runout is accounted for and corrected by a pupil tip-tilt mirror at the entrance of the instrument. The near-IR Strehl ratio is $>$95~\%. In particular, non-common path aberrations (from the DM to the detector) are measured off-line using a phase diversity algorithm and compensated by reference slopes adjustment. In these conditions, static wavefront errors left uncorrected in the system are estimated at the level of $\sim$$6$ nm rms. 

SPHERE offers various coronagraphic possibilities (classical Lyot coronagraphs, apodized Lyot coronagraphs -- ALC --, and achromatic four quadrants phase masks). During the experiment, a 5.2$\lambda/D$ ALC has been used. 

The infra-red dual beam imaging and spectroscopy (IRDIS) sub-system includes a spectral range from 950 to 2320 nm and an image scale of 12.25 mas per pixel (Nyquist sampling at 950nm). The field of view is greater than 11$\arcsec$ square, with a 2kx2k Hawaii-II-RG detector. The main mode of IRDIS is the dual band imaging (DBI), providing images in two neighboring spectral channels with minimized differential aberrations. 
Ten different filter couples are defined corresponding to different spectral features in modeled exoplanet spectra. During the experiment the narrow $H$-band couple filters have been used (centered around 1593 nm, and 1667 nm, R=30). The DBI mode is not of the interest of the present study as we are interested in the temporal evolution of aberrations, nevertheless it allows a simulatenous characterization of quasi-static speckles at two different wavelengths for comparison/confirmation purpose. \\
Finally, the SPHERE instrument is installed on an active tripod damping system, and fully covered by an hermetic enclosure. 

\subsection{Thermal functioning}
The SPHERE instrument must conform to several environmental specifications. In particular, it shall operate under a temperature range from 5 to 18$^{\circ}$C, while the highest gradient of temperature at the VLT is -0.9$^{\circ}$C/h and -1.4$^{\circ}$C/h for respectively 80$\%$ and 95$\%$ of the nights.
In order to validate that SPHERE is compliant with such conditions, cold tests have been set up for functional and performance evaluation at different ambient temperatures within the operational range, and with realistic transient conditions reflecting situations encountered at the VLT. The whole SPHERE instrument has been installed in a cold tent (150 m$^{3}$) and cooled down with an efficient air conditioning system.
The cooling system allows to reduce by $\sim$10$^{\circ}$C the temperature inside the tent compared to the outside (in the integration hall). Numerous temperature probes have been installed at several critical locations in the instrument to accurately monitored  the evolution inside/outside the SPHERE enclosure.

\subsection{Observing run and data reduction}
The observing run consists in recording a time series of AO close-loop IRDIS coronagraphic images. Each coronagraphic image of the time series corresponds to a series of 3~s short exposure images, averaged out over $\sim$3~mn. During the experiment, the temperature outside the enclosure (inside the cold tent) was monitored and programmed to decrease from 15$^{\circ}$C with a rate of 2$^{\circ}$C per hour.  
The total duration of the experiment was 100~mn. The gradient inside the enclosure was estimated at the level of 0.2$^{\circ}$C/h.  
The data-reduction process corrects for bad pixels and background, and normalizes the images with respect to exposure time and flux. Before subtraction, a fine, sub-pixel correction of the residual tip-tilt component is performed on the raw images. The analysis has been done on the two $H$-band filters data set available and converge to similar results, so that only one data set is presented here. 
Depending on the nature of the image analyzed, we applied different metrics.
The \textit{contrast} refers to the ratio of intensity in the raw coronagraphic image, averaged azimuthally at a given angular separation, to the peak intensity of the direct flux. 
When studying a differential image, implying the subtraction of a reference image, the average contrast is no longer suited.
The \textit{detectability} is used then, which stands for the azimuthal standard deviation measured in a ring of width $\lambda/D$. It quantifies the ability to distinguish a companion at a given angular distance.
\begin{figure*}[ht!]
\centering
\includegraphics[width=5.6cm]{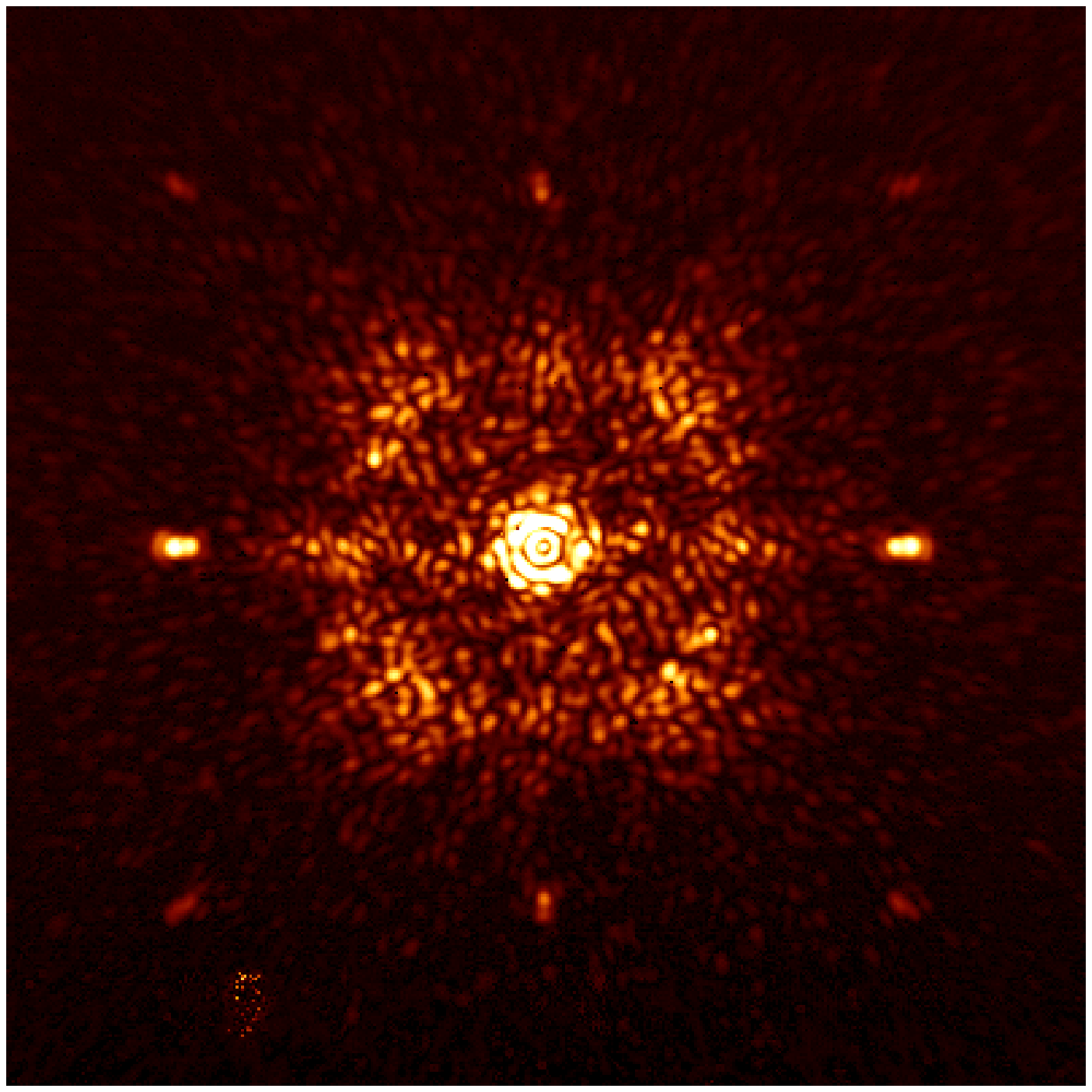}
\includegraphics[width=5.6cm]{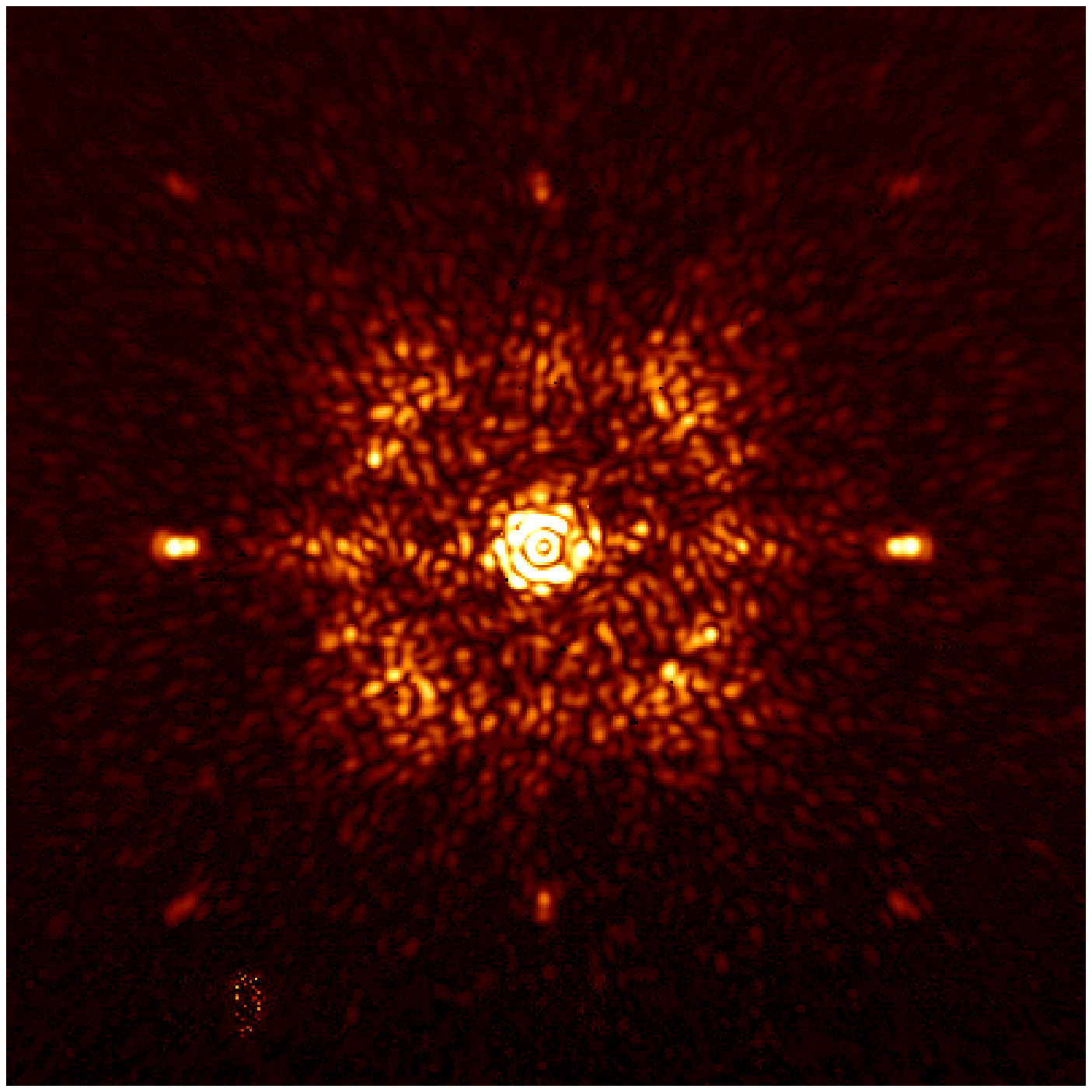}
\includegraphics[width=5.6cm]{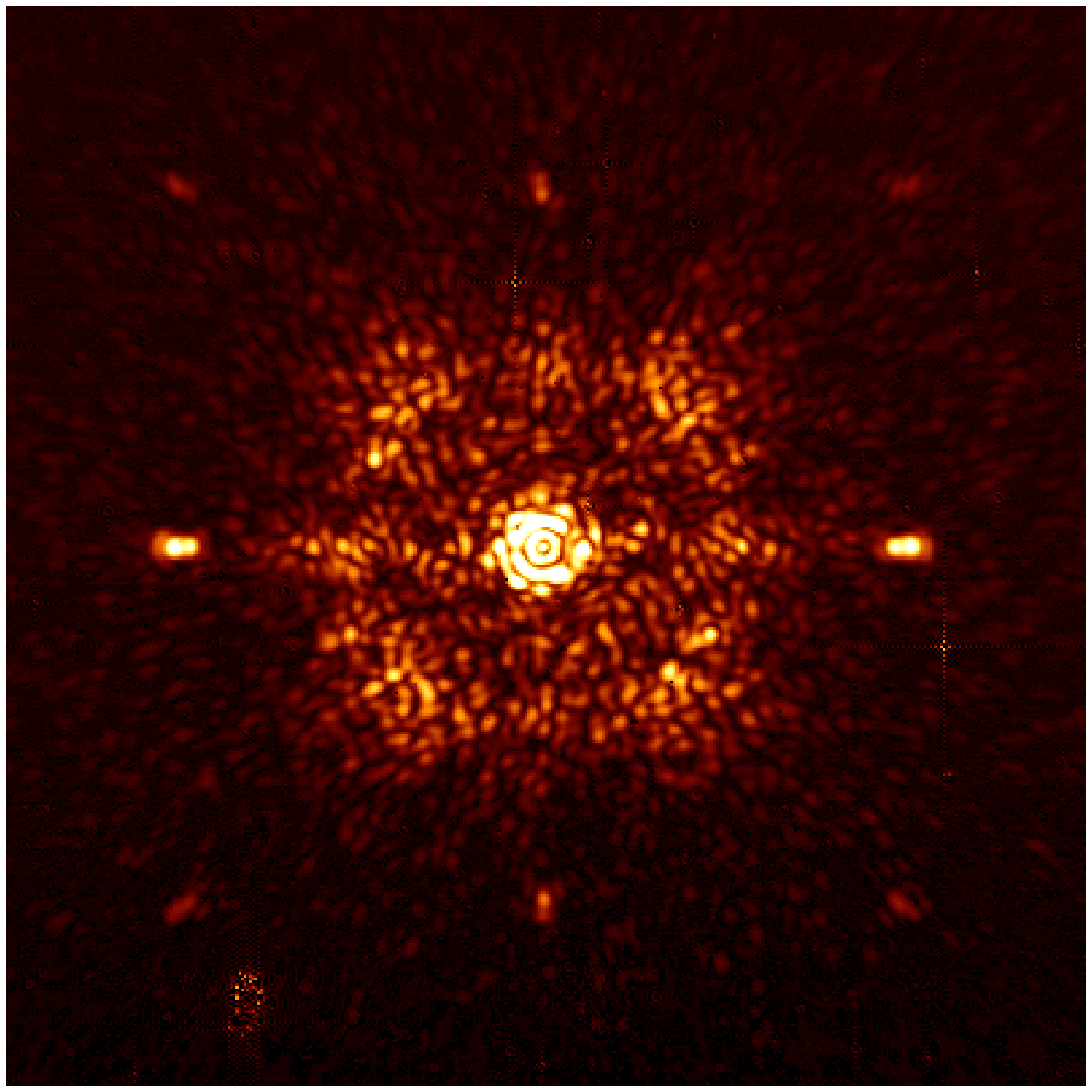}
\caption{Coronagraphic images recorded at $t_{0}$, $t_{0} + 10$~mn, and $t_{0} + 100$~mn. The Strehl ratio is  $\sim$~95~\%. The arbitrary color scale and dynamic range (identical for the three images) were chosen to enhance the contrast for the sake of clarity. The field covered in the images is $\sim$ 2$\arcsec$ square.} 
\label{Images}
\end{figure*} 
\begin{figure*}[ht!]
\centering
\includegraphics[width=8.5cm]{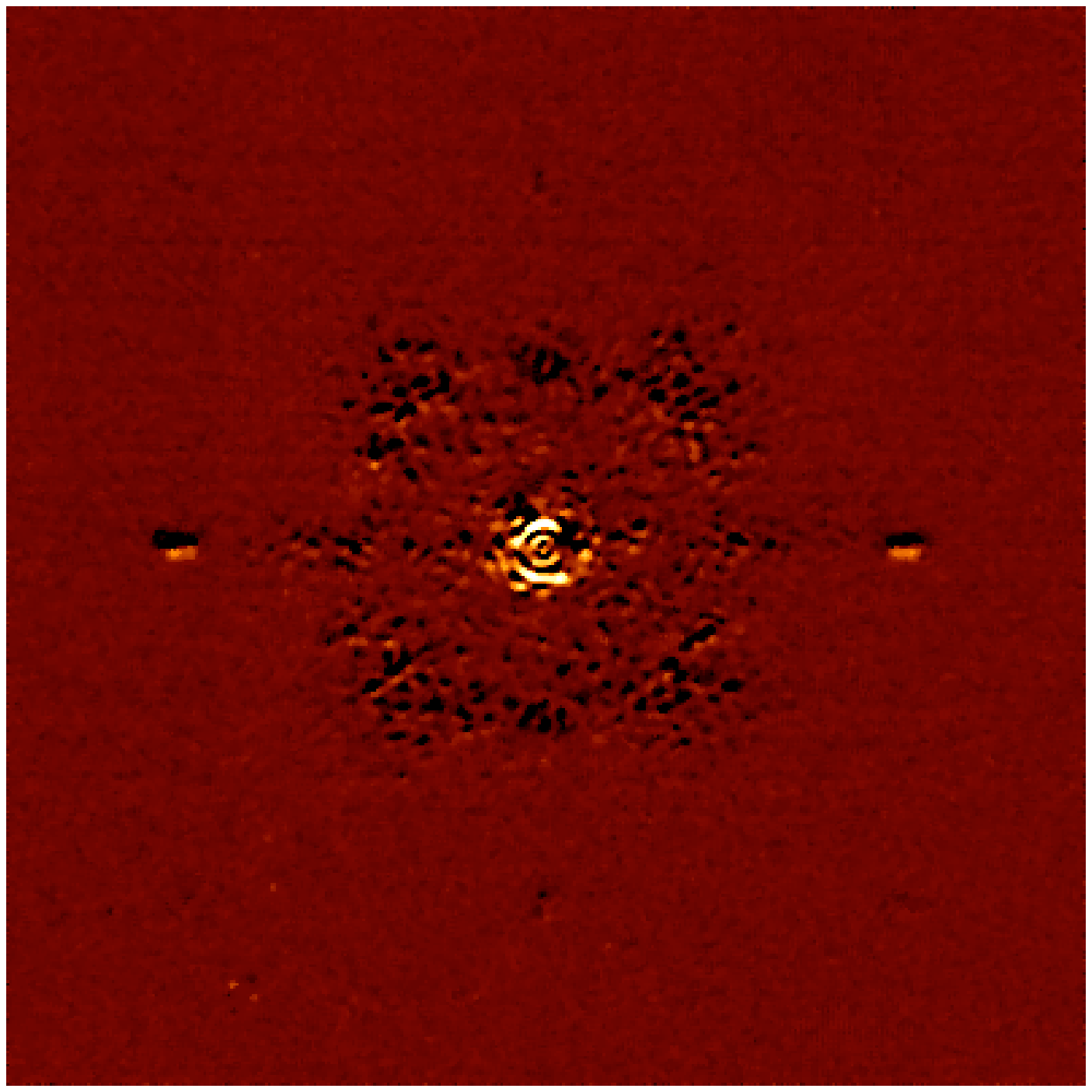}
\includegraphics[width=8.5cm]{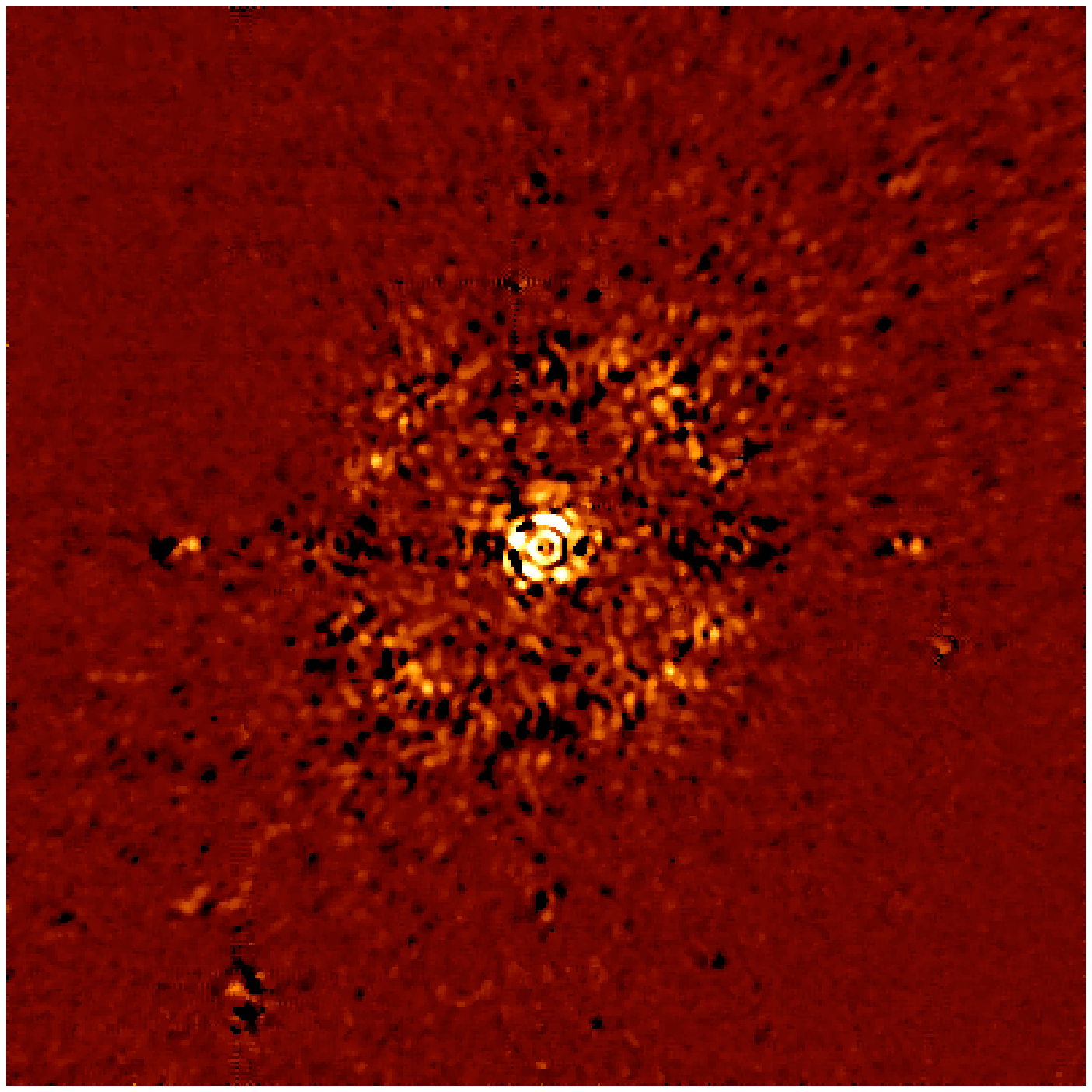}
\caption{Differential coronagraphic images. Left: difference of the $t_{0} + 10$~mn image to the reference $t_{0}$. Right: difference of the $t_{0} + 100$~mn image to the reference $t_{0}$. The increase in the strength of residuals can be qualitatively observed here (the intensity range is similar in both images).}
\label{Images2}
\end{figure*} 
\begin{figure}[ht!]
\centering
\includegraphics[width=8.5cm]{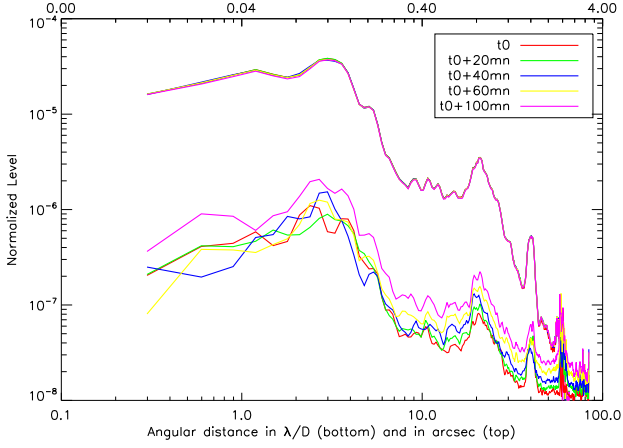}
\caption{Contrast profiles of a time series of coronagraphic images (top) and detectability (1$\sigma$) of the differential images (subtraction of the time series of coronagraphic  images to the reference one, bottom)}
\label{res}
\end{figure} 

\section{Analysis and interpretation}
Figure \ref{Images} presents three coronagraphic images extracted from the time series: the coronagraphic image recorded at $t_0$ (the reference), $t_0$+10mn, and $t_0$+100mn. 
In Fig.~\ref{Images2}, the corresponding differential images are shown: the difference of the $t_{0} + 10$~mn image to the reference $t_{0}$ (left), and the difference of the $t_{0} + 100$~mn image to the reference $t_{0}$ (right). The corresponding profiles for both raw and differential images are presented in Fig.~\ref{res}.\\
Qualitatively, the raw coronagraphic images (Fig.~\ref{Images}) demonstrate starlight attenuation, and exhibit static speckles with lower intensity in the AO-correction domain (from the second or third wing of the coronagraphic image to the rise of the AO cut-off frequency).  A radial trend in speckle intensity is observable in the image: speckles closer to the center of the image are brighter.
The central part of the coronagraphic image is dominated by diffraction residuals. 
The AO cut-off frequency  can be readily seen in the image owing to the slope of intensity in the speckle field at 0.8$\arcsec$ (20$\lambda$/D) from the center, as expected. 
Outside the inner-domain defined by the AO cut-off frequency, the AO system cannot measure or correct the corresponding spatial frequencies. Various bright spots are observable at $\sim$40$\lambda/D$ (twice the AO cut-off frequency) and correspond to the inter-actuator pitch spatial frequency. 

As observed in Fig.~\ref{Images} and presented in Fig.~\ref{res} (top curves), raw coronagraphic images are dominated by the static contribution, for which contrast profiles are stable over time at any angular separation. 
Raw coronagraphic images are dominated by static speckle noise, as no evolution in the speckle field is visually detectable over the three images presented that cover the temporal duration of the experiment.
This is consistent with the fact that the interaction between the quasi-static terms of Eq.~\ref{A}, being time-dependent, and static terms, assumed time-independent, can be studied through differential imaging from a time series of raw coronagraphic images, and not directly at the level of raw images.

The static wavefront error amplitude has been evaluated on the raw coronagraphic images using Eq.~\ref{F}, and converges to the value of $\sim$5~nm~rms, though it might be slightly underestimated (see Sect. \ref{refStatic}). 
This is basically a factor of 10 lower than the estimation obtained on the HOT bench, and contribute to the increase of dynamic range of the instrument to that of the performance obtained with the HOT bench in both classical and differential imaging (virtually by an order of magnitude). 
By contrast to raw images, an evolution in the speckle field is seen in the differential images. 
\begin{figure}[ht!]
\centering
\includegraphics[width=8.5cm]{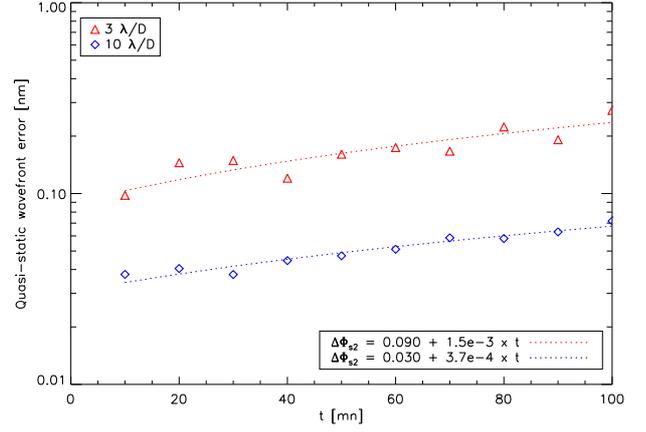}
\caption{Time variability of wavefront error due to quasi-static speckles, evaluated at various angular separations (observational data).}
\label{res2}
\end{figure} 
The profiles presented in Fig.~\ref{res} (bottom curves) clearly indicate that the detectability level degrades with time. These profiles demonstrate that raw coronagraphic images evolve temporally, being less and less correlated with the reference over time, even though such an evolution cannot be readily seen in raw images. 
Further, this degradation of the detectability is effective at all angular separations. 
This result is compliant with observations formally reported in Paper I. 
From the time series of differential images, using Eq.~\ref{E}, we derive the quasi-static wavefront error contribution per Fourier component, of the pinning effect at several angular separations. 
Figure~\ref{res2} shows the temporal evolution of $\phi_{s2}$ at 3, and 10~$\lambda/D$, i.e.\ the quasi-static wavefront error (rms) as function of time at several angular separations. It clearly indicates that  $\phi_{s2}$ is time-dependent and increases with time, justifying the constant degradation of detectability observed in Fig.~\ref{res} (bottom curves). This is true for all angular separations, and the shorter the separation the higher the amplitude. 
Quantitatively, the level of quasi-static wavefront error that limits the sensitivity in the differential images ($\sim$$10^{-7}$ to $\sim$$10^{-8}$, 1$\sigma$) is found to be in the regime of $\sim$ 0.1$\AA$ to 1$\AA$. 
The power law of the temporal evolution of the quasi-static wavefront error is derived and exhibits a similar tendency whatever the angular position in the field. 
It can be fitted by a linear function of time. The parameters for the linear fits are presented in the legend of  Fig.~\ref{res2}, and can be used to extrapolate a model for quasi-static speckle evolution in the context of high-contrast imaging instruments. From the linear fits derived at various angular separations, and considering that the difference in amplitude is not significant , we attempt to generalize the expression of the power law for any angular position in the field, which reads as the following approximation:
\begin{equation}
\Delta \phi_{s2} (t) \simeq 0.065 + 0.001 \times t,
\label{FIT}
\end{equation}
\noindent where $t$ is the time in minutes, and $\Delta \phi_{s2}$ is expressed in nm~rms.

In Paper I, the parameters found for the fit were 0.250 for the value at the origin, and a slope of 0.012. 
This means that SPHERE is definitely more stable in term of thermo-mechanical variations than the HOT bench, though this was expected (a slope reduced by at least an order of magnitude, since data were not recorded a transient regime as performed here, but in stabilized temperature environment).   
While scaling the fit parameters to model quasi-static temporal evolution observed with the HOT bench to real instrument was highly non-trivial as it significantly dependents on the instrument environment (temperature or pressure changes, mechanical flexures, etc...), we believe that the estimates found with SPHERE and presented in Eq.~\ref{FIT} are representative of the new high-contrast imaging instrument generation that SPHERE represents, and can be used as such. In our data we found that quasi-static wavefront error increases with $\sim$0.7$\AA$ per minute. 
\begin{figure}[ht!]
\centering
\includegraphics[width=8.5cm]{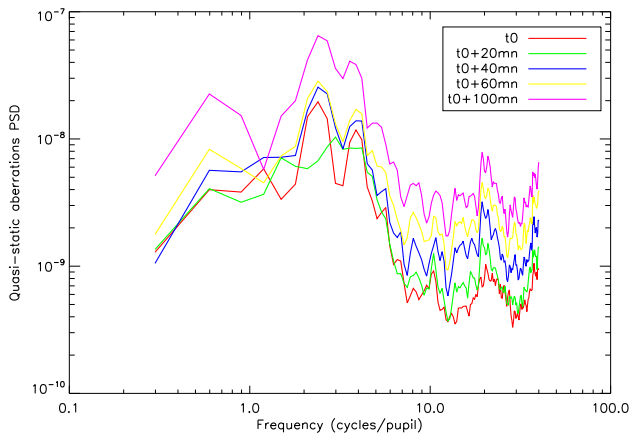}
\caption{Quasi-static aberrations power spectral density at various timescales (using Eq. 5).}
\label{res3}
\end{figure} 
Quasi-static speckles, fast-evolving on the level of a few angstroms over a timescale of few seconds, explained the evolution of our sensitivity through amplification of the systematics. It is believed that this effect is a consequence of thermal and mechanical instabilities (pressure changes, mechanical flexures...) of the optical bench. This is qualitatively supported by the fact that outside the AO control radius, a characteristic "butterfly shape" drawn by the speckle pattern in the differential image at $t_0+100mn$ is observable (Fig. \ref{Images2}, right). This indicates that a beam-shift is at work, though it is difficult to obtained quantitative information on it.

$I$ functions described in Sect. 2 essentially represent power spectral densities (PSD). While $I_{c}$  symbolizes the PSD of the static wavefront, $\Delta I_{s2}$ stands for the PSD of the differential aberrations, which is calculated by Eq. \ref{C}. Plotting Eq. \ref{C} gives access to the quasi-static aberration PSD as function of time (Fig. \ref{res3}). 
From the DSPs presented in Fig. \ref{res3}, we can observe that:
(1/) at very low frequencies (from 0.1 to 3 cycles per pupil) the PSD is temporally roughly stable and exhibits a $f^{0}$ power law, where $f$ is the spatial frequency, while it must be pointed out that in this frequency domain, Eq. \ref{C} might not be entirely valid, 
(2/) at low frequencies (from 3 to 8 cycles per pupil) the PSD exhibits a $f^{-5}$ power law, (3/) at intermediate frequencies (from 8 to 20 cycles per pupil) the PSD exhibits a $f^{0}$ power law, then essentially dominated by a white noise, (4/) at high frequencies (outside the AO control domain, from 20 to 30 cycles per pupil, noise dominated at farther spatial frequencies) the power law is again $f^{-5}$. 
From Fig. \ref{res3} it is difficult to extract further unambiguous informations, or initiate preliminary explanations, such as the $f^{-5}$ power law being in contradiction with the generally adopted and standard PSD slope of $f^{-2}$ for surface optics (static aberrations).  

\section{Conclusion}
In this paper, we confirm the results formerly reported in Paper I with new measurements, and derive a practical and generalized expression to model quasi-static speckles temporal evolution for any angular position in the field (Eq.\ref{FIT}), in the context of the forthcoming high-contrast planet imagers. 
Quasi-static aberrations observed in a time series of extreme adaptive optics-corrected coronagraphic images exhibit a linear power law with time and are interacting through the pinning effect with static aberrations. 
We examine and discuss this effect using the statistical model of the noise variance in high-contrast imaging proposed by \citet{Soummer07}, where the effect of pinning quasi-static to static speckles as described by the expression for the variance (Eq.~\ref{E}) is found to reflect the situation in our data set fairly well.

We found that quasi-static wavefront error increases with a rate of $\sim$0.7$\AA$ per minute.
In this context, Eq. \ref{FIT} is meant to feed complex system analysis for timescale specifications and impact assessments in the context of calibration/operational strategies for high-contrast imaging class instruments. 
In addition, Eq. \ref{C} provides a useful insight in the PSD of quasi-static aberrations in the system, and at first glance to identify what moves in the system. 
The proposed model of quasi-static speckles temporal evolution can be used to derive timescales for calibration/operational aspects, such as ADI, or non-common path aberrations correction. 

The case considered in this present paper represents a static system subject to a thermal gradient. The foreseen impact of rotating components in the instrument, such as, e.g., the atmospheric dispersion compensators (ADCs, not seen by the AO system), or the derotator (seen by AO system) will be treated in a separated study.

\acknowledgements
SPHERE is an instrument designed and built by a consortium consisting of IPAG, MPIA, LAM, LESIA, Laboratoire Lagrange, INAF, Observatoire de Gen\`{e}ve, ETH, NOVA, ONERA, and ASTRON in collaboration with ESO. 

\bibliography{MyBiblio} 

\end{document}